\documentclass[aps,prd,preprint,groupedaddress]{revtex4}

\usepackage{graphicx}
\usepackage{amsmath}
\usepackage{bm}

\def\ee{\end{eqnarray}}

\def\=:{=\hspace{-.7em}\raisebox{1.1ex}{.}\hspace{.1em}\raisebox{-0.2ex}{.} }

\newcommand {\beq}{\begin{eqnarray}}
\newcommand {\eeq}{\end{eqnarray}}
\newcommand {\non}{\nonumber\\}

\newcommand {\1}[1]{\frac{1}{#1}}

\newcommand {\ph}{\varphi}

\newcommand {\tr}{{\rm tr}\,}

\newcommand{\hs}[1]{\hspace{#1 mm}}

\begin{document}


\title{Defect formation from 
defect--anti-defect annihilations
}


\author{Muneto Nitta$^{1}$}

\affiliation{
$^1$Department of Physics, and Research and Education Center for Natural 
Sciences, Keio University, Hiyoshi 4-1-1, Yokohama, Kanagawa 223-8521, Japan\\
}


\date{\today}
\begin{abstract}
We show that when a topological defect 
with extended world-volume annihilates with an anti-defect, 
there arise topological defects with 
dimensions less than those of the original defects by one. 
Domain wall annihilations create vortices while 
monopole-string annihilations result in instantons. 
We find that twisted domain wall rings are vortices, 
whereas twisted monopole rings are instantons.

\end{abstract}

\maketitle

\section{Introduction}

When a particle and its anti-particle meet, 
they annihilate and turn to energy or 
decay into other particles. For instance,
an electron and a positron on collision annihilate 
each other to create a photon. 
Similarly, when a soliton and an anti-soliton collide, 
they also annihilate and usually decay into 
elementary excitations. 
For instance, in superfluids, 
a vortex and an anti-vortex pair annihilate on collision 
and decay into phonons. 
A question that arises along similar line is what happens 
if an extended object and its counter object collide. 
Some years ago, 
this was studied in string theory 
for a Dirichlet(D) $p$-brane, 
which is an extended object (soliton) 
with a world-volume with a $p$-dimension.
It has been found that 
the annihilation of a D$p$-brane with an anti-D$p$-brane 
results in the creation of D$(p-2)$-branes \cite{Sen:2004nf}. 
Then, the question that arises is whether such a process 
exists for solitons as extended objects in field theory. 

In this paper, we show that when 
a soliton and an anti-soliton collide and annihilate,  
there arise topological defects with 
dimensions less than those of the original defects by one. 
This is inevitable when topological defects have 
a $U(1)$ modulus (collective coordinate).
The examples are 
1) when a domain wall and an anti-domain wall annihilate 
in 2+1 dimensions, 
there appear vortices, while  
2) when a monopole string and anti-monopole string 
annihilate in 4+1 dimensions, 
there appear Yang-Mills instantons.
As a byproduct of our results, 
we find that twisted domain wall (monopole) rings are 
equivalent to vortices (instantons).


\begin{figure}[h]
\begin{center}

\includegraphics[width=0.2\linewidth,keepaspectratio]{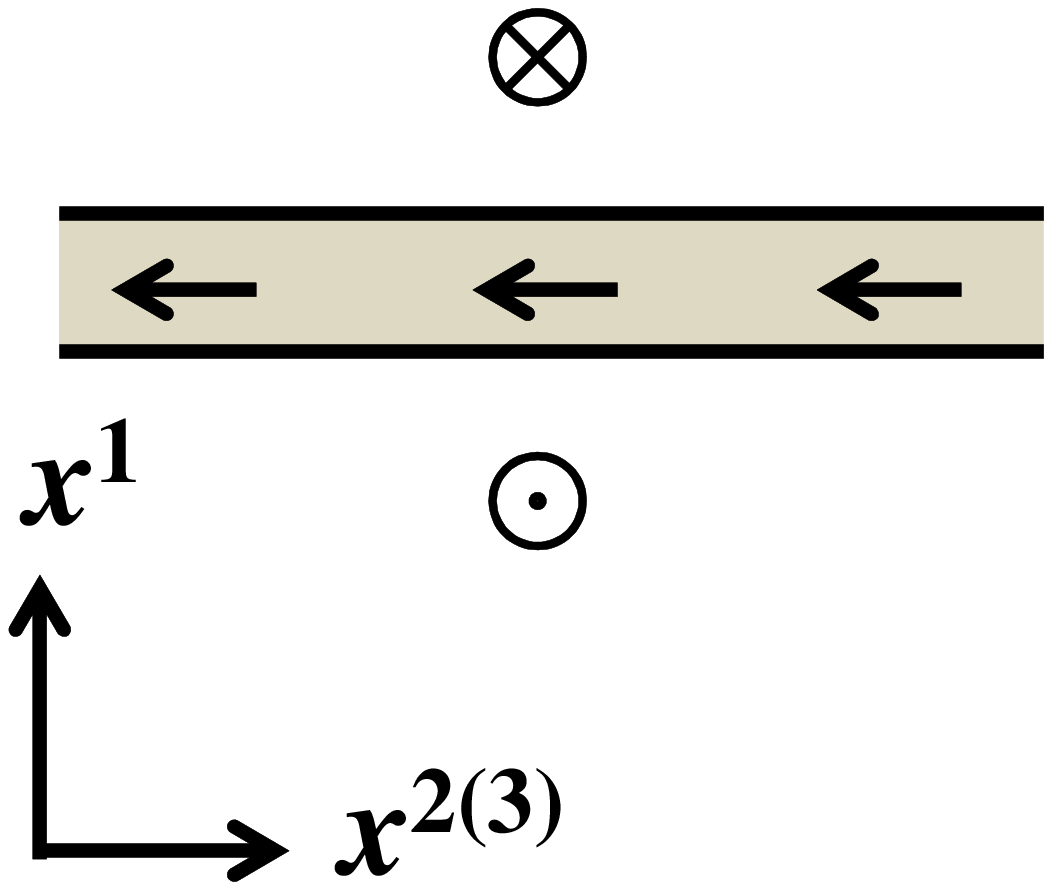}
\quad
\includegraphics[width=0.2\linewidth,keepaspectratio]{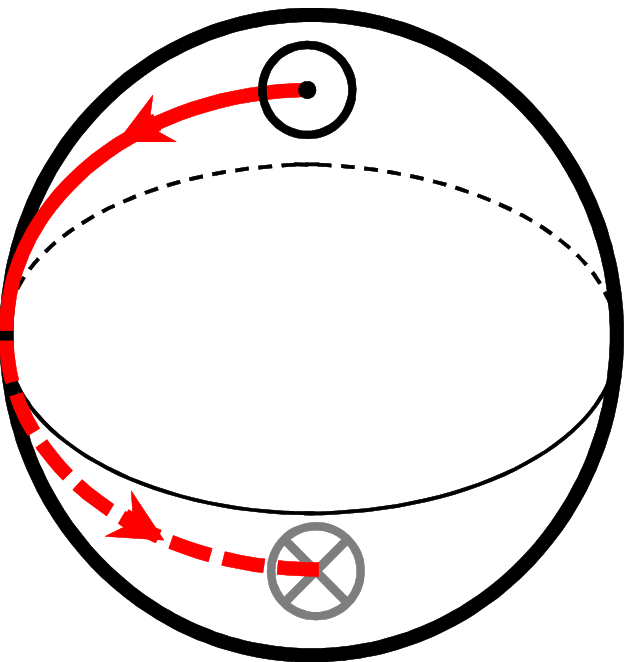}
\quad
\includegraphics[width=0.2\linewidth,keepaspectratio]{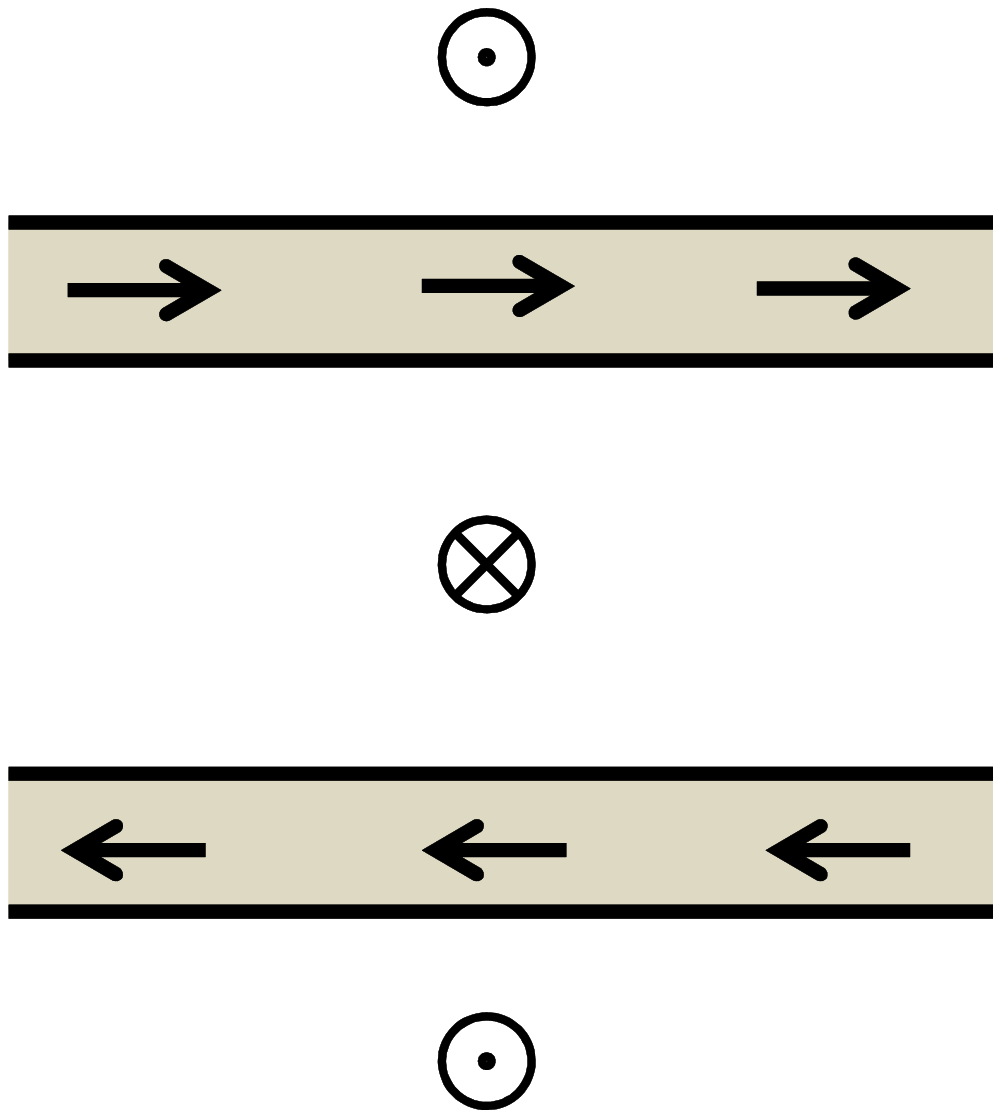}
\quad
\includegraphics[width=0.2\linewidth,keepaspectratio]{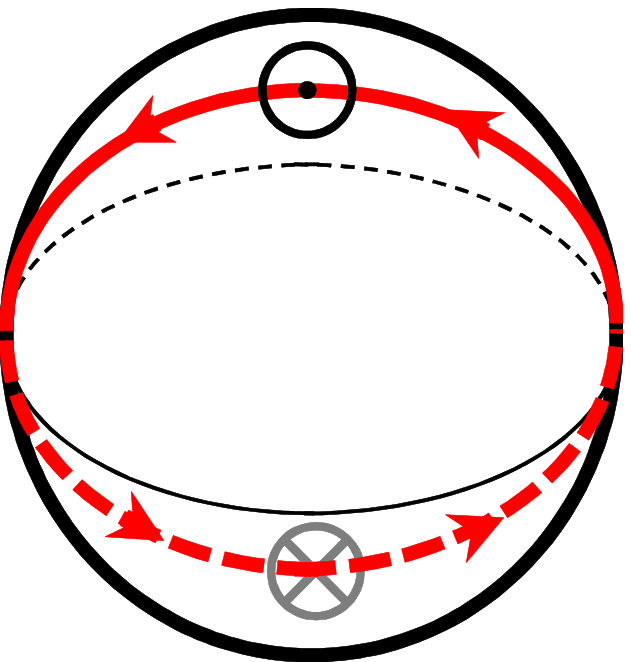}\\
(a)\hspace{3.5cm} (b)\hspace{3.5cm}
(c)\hspace{3.5cm} (d)

\caption{(a,b) Single domain wall and (c,d) a pair of a domain wall and 
an anti-domain wall in the ${\bf C}P^1$ model. 
(a) A single domain wall.
The wall is perpendicular to the $x^1$-axis.
The arrows denote points in the ${\bf C}P^1$.
(b) The ${\bf C}P^1$ target space. 
The north and south poles are denoted by 
$\odot$ and $\otimes$, respectively.
The path connecting them represents the map from 
the path in (a) along the $x^1$-axis 
in the real space from $x^1 \to - \infty$ to $x^1 \to + \infty$. 
The path in the ${\bf C}P^1$ target space 
passes through one point on the equator, 
which is represented by ``$\leftarrow$" in (a) in this example. 
In general, the $U(1)$ zero modes are localized on the wall.
\\ 
(c) The wall and anti-wall 
configuration. 
The $U(1)$ zero modes on the wall and anti-wall
are opposite to each other. 
(d) The path represents the map from the wall and anti-wall 
configuration. 
}
\label{fig:brane-anti-brane-2d} 
\end{center}
\end{figure}

\section{Vortices from Wall--anti-wall Annihilation}
As the simplest example, we first consider 
a domain wall pair annihilation. 
Domain walls are extended objects in $d=2+1$ dimensions or more.
We consider the ${\bf C}P^1$ model, 
also known as the $O(3)$ model, 
which describes ferromagnets. 
In order to admit domain walls, 
we consider the potential term 
admitting two minima, 
which gives a one-axis isotropy in the context of the magnetism.
The Lagrangian is given by 
\beq
 {\cal L} = \1{2} \partial_{\mu}{\bf n} \cdot \partial^{\mu}{\bf n} 
 - m^2(1-n_3^2) , 
 \quad {\bf n}^2 =1.
\eeq 
with a three-vector of scalar fields,  
${\bf n} = (n_1(x),n_2(x),n_3(x))$. 
This model has two discrete vacua, 
$n_z=\pm 1$, corresponding to the north and south poles 
of the sphere.
It is also known as 
the massive ${\bf C}P^1$ model \cite{Abraham:1992vb}, 
which shows supersymmetry when fermions are suitable 
introduced in it,  
although supersymmetry 
is not essential in our study. 
It is convenient to use the ${\bf C}P^1$ projective 
coordinate $u$, 
considering which the Lagrangian can be rewritten as \cite{Abraham:1992vb}
\beq
&& {\cal L} =
 {\partial_{\mu} u^* \partial^{\mu} u - m^2 |u|^2 
  \over (1 + |u|^2)^2}  
\label{eq:Lagrangian}
\eeq
with ${\bf n} = \Phi^\dagger {\bf \sigma} \Phi$, 
where $\Phi^T = (1,u)/\sqrt{1 + |u|^2}$.
The two vacua, $n_3=+1$ and $n_3=-1$, are mapped to $u=0$ and $u=\infty$, 
respectively.

There exists a domain wall interpolating 
the two discrete vacua \cite{Abraham:1992vb}
\beq
 u_{dw} = e^{m (x^1-x^1_0) + i \ph}  \label{eq:wall-sol}
\eeq
with width $1/m$ and tension $m$.
Here, $x^1_0$ and $\ph$ are real constants called 
moduli parameters or collective coordinates.
They are Nambu-Goldstone modes associated with 
broken translational and internal $U(1)$ symmetries, 
respectively.
A domain wall is mapped to 
a large circle starting from the north pole, denoted by $\odot$,
and ending up at the south pole, denoted by $\otimes$,
in the ${\bf C}P^1$ target space 
(Fig.~\ref{fig:brane-anti-brane-2d}~(b)).
An anti-domain wall is obtained simply by introducing 
a minus sign in front of 
$x^1-x^1_0$ in Eq.~(\ref{eq:wall-sol}).

\begin{figure}
\begin{center}
\includegraphics[width=0.2\linewidth,keepaspectratio]{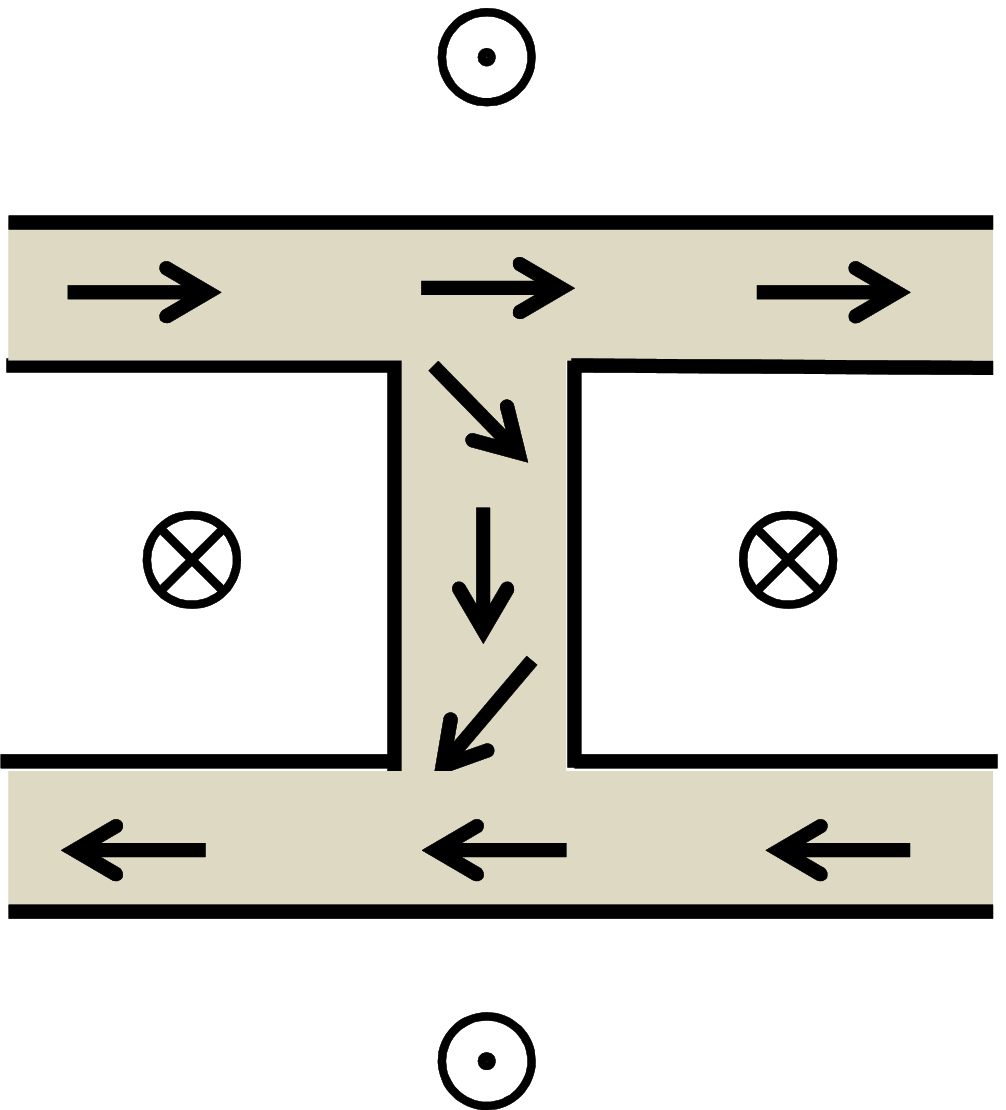}
\quad
\includegraphics[width=0.2\linewidth,keepaspectratio]{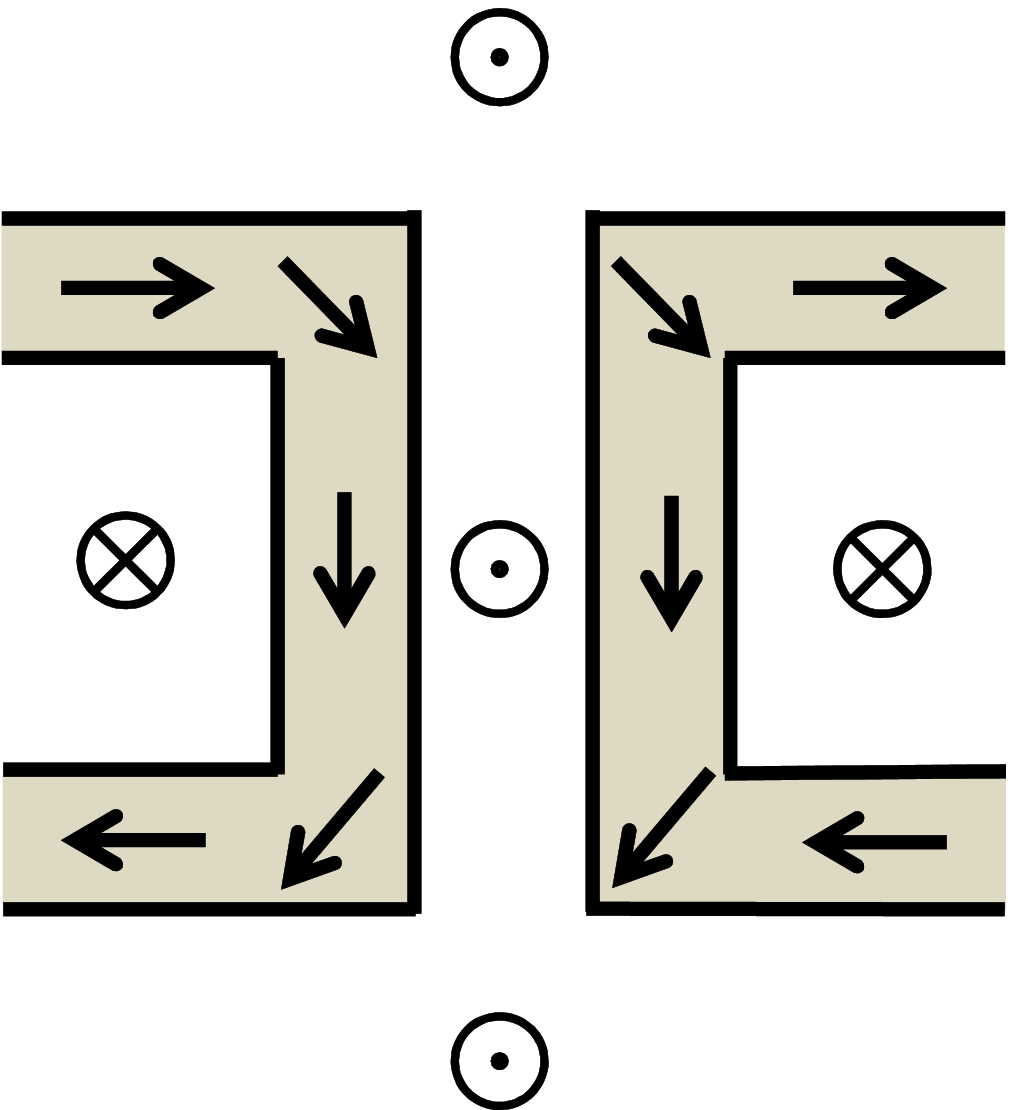}
\quad
\includegraphics[width=0.2\linewidth,keepaspectratio]{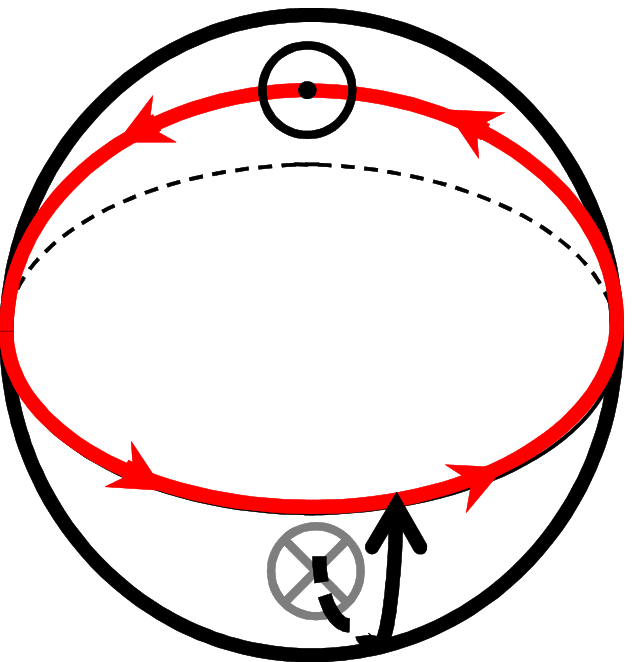}
\\
(a)\hspace{3cm} (b)\hspace{3cm} (c)\\

\medskip
\includegraphics[width=0.2\linewidth,keepaspectratio]{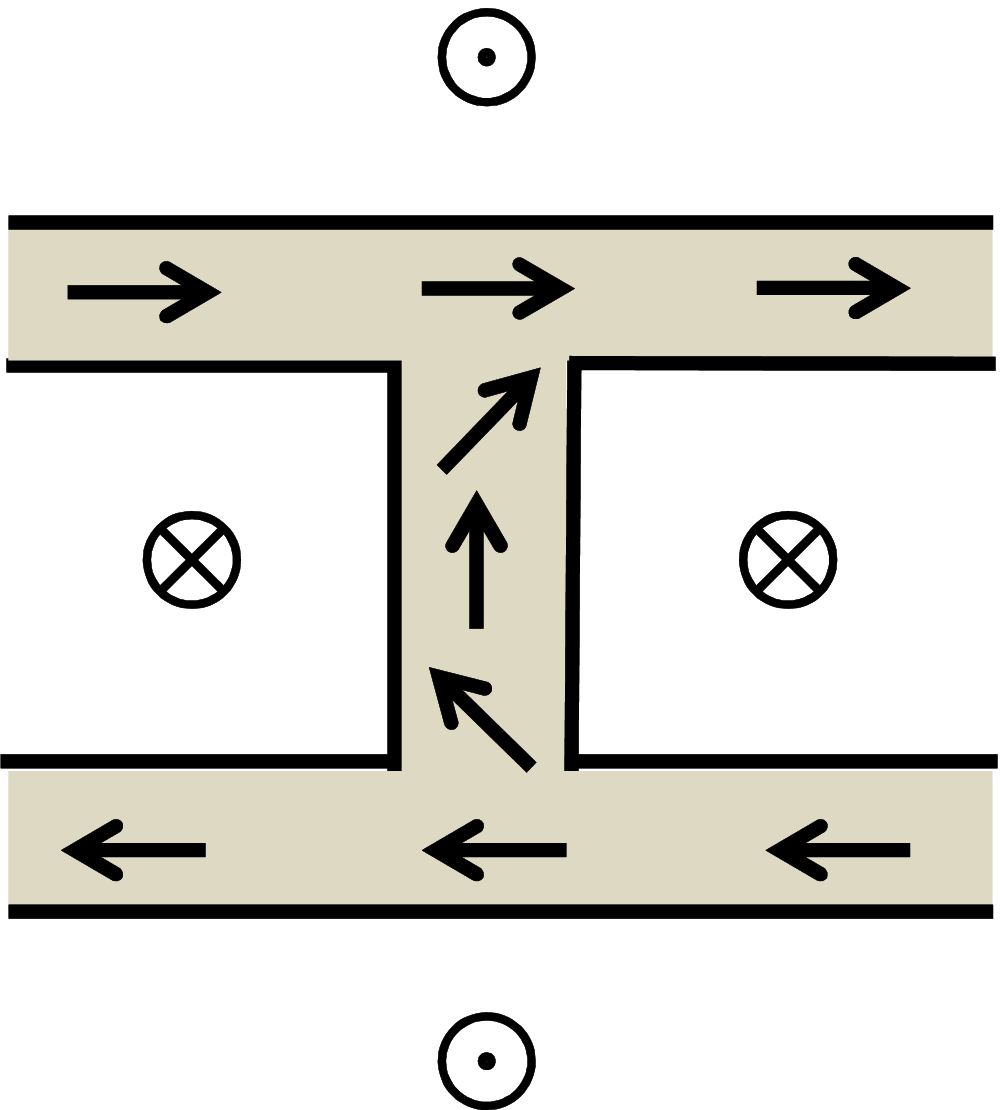}
\quad
\includegraphics[width=0.2\linewidth,keepaspectratio]{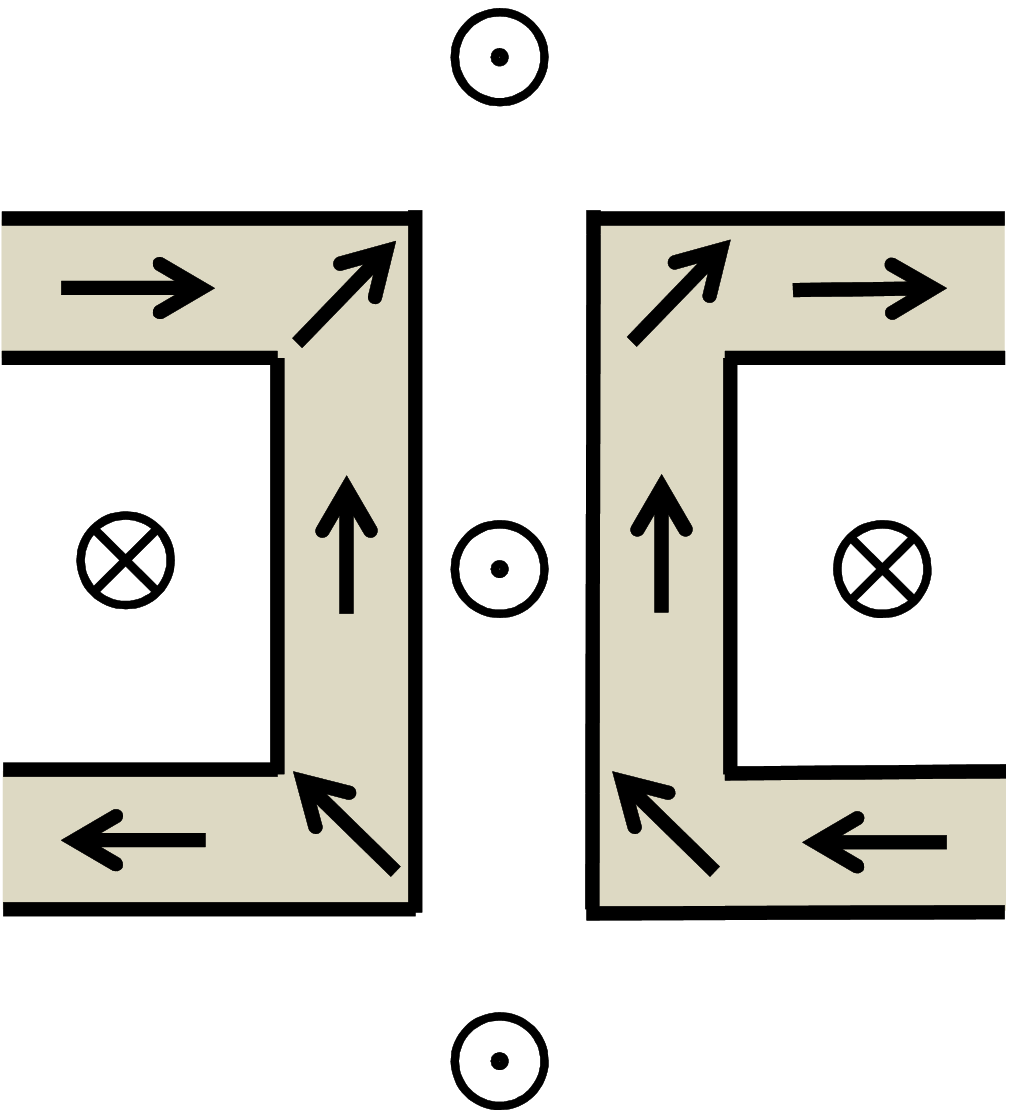}
\quad
\includegraphics[width=0.2\linewidth,keepaspectratio]{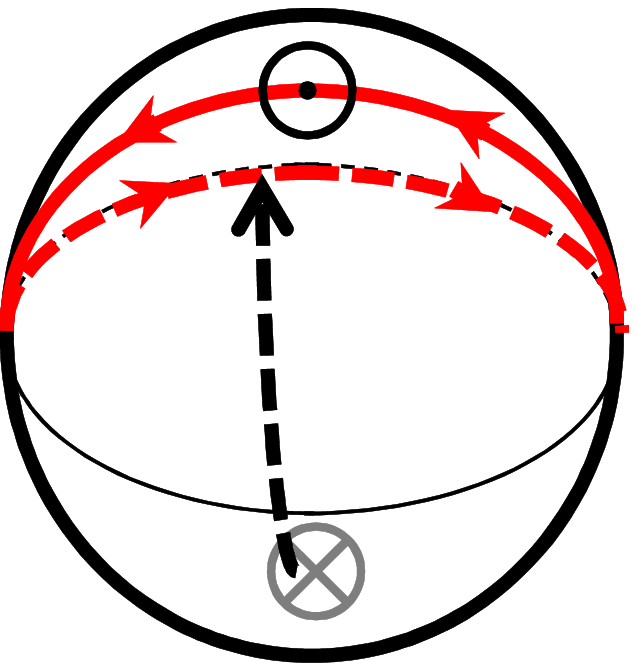}
\\
(d)\hspace{3cm} (e)\hspace{3cm} (f)
\caption{
Decaying processes of the wall and anti-wall.
(a,d) A bridge is created between the wall and the anti-wall. 
In this process, there are two possibilities of 
the ${\bf C}P^1$ structure 
along the bridge. 
(b,e) The upper and lower regions are connected 
by breaking the bridge. 
(c,f) Accordingly, the loop in the ${\bf C}P^1$ target space 
is unwound in two ways.
\label{fig:wall-anti-wall-annihilation} 
} 
\end{center}
\end{figure}

Next, we consider a domain wall and an anti-domain wall.
Here, the $U(1)$ zero modes of the wall and 
anti-wall are considered to be opposite, 
as in Fig.~\ref{fig:brane-anti-brane-2d}~(c).
The configuration is mapped to a loop 
in the ${\bf C}P^1$ target space 
(Fig.~\ref{fig:brane-anti-brane-2d}~(d)).
However, this configuration is unstable  
as it should end up with the vacuum with the up-spin $\odot$.
In the decaying process, the loop is unwound from 
the south pole in the target space. 
The unwinding of the loop can be achieved in 
two topologically inequivalent ways, 
which are schematically shown in Fig.~\ref{fig:wall-anti-wall-annihilation} 
(c) and (f). 
In the real space, 
at first, a bridge connecting the two walls is created 
as in Fig.~\ref{fig:wall-anti-wall-annihilation} (a) and (d).
Here, there exist two possibilities of the spin structure of the bridge, 
corresponding to two ways of the unwinding processes. 
Along the bridge in the $x^1$-direction, 
the spin rotates (a) anti-clockwise or (b) clockwise on the equator 
of the ${\bf C}P^1$ target space.
Let us label these two kinds of bridges by 
``$\downarrow$" and ``$\uparrow$", respectively.

In the next step, the bridge is broken into two pieces, 
as in Fig.~\ref{fig:wall-anti-wall-annihilation} (a) and (d), 
with the vacuum state, i.e., the up-spin $\odot$ state, 
filled between them. 
Let us again label these two kinds of holes by 
``$\downarrow$" and ``$\uparrow$", respectively. 
In either case, the two regions separated by the domain walls 
are connected through a hole created by the decay of the domain walls.
Once created, these holes grow, thus reducing the wall tension.
\begin{figure}
\begin{center}
\includegraphics[width=0.2\linewidth,keepaspectratio]{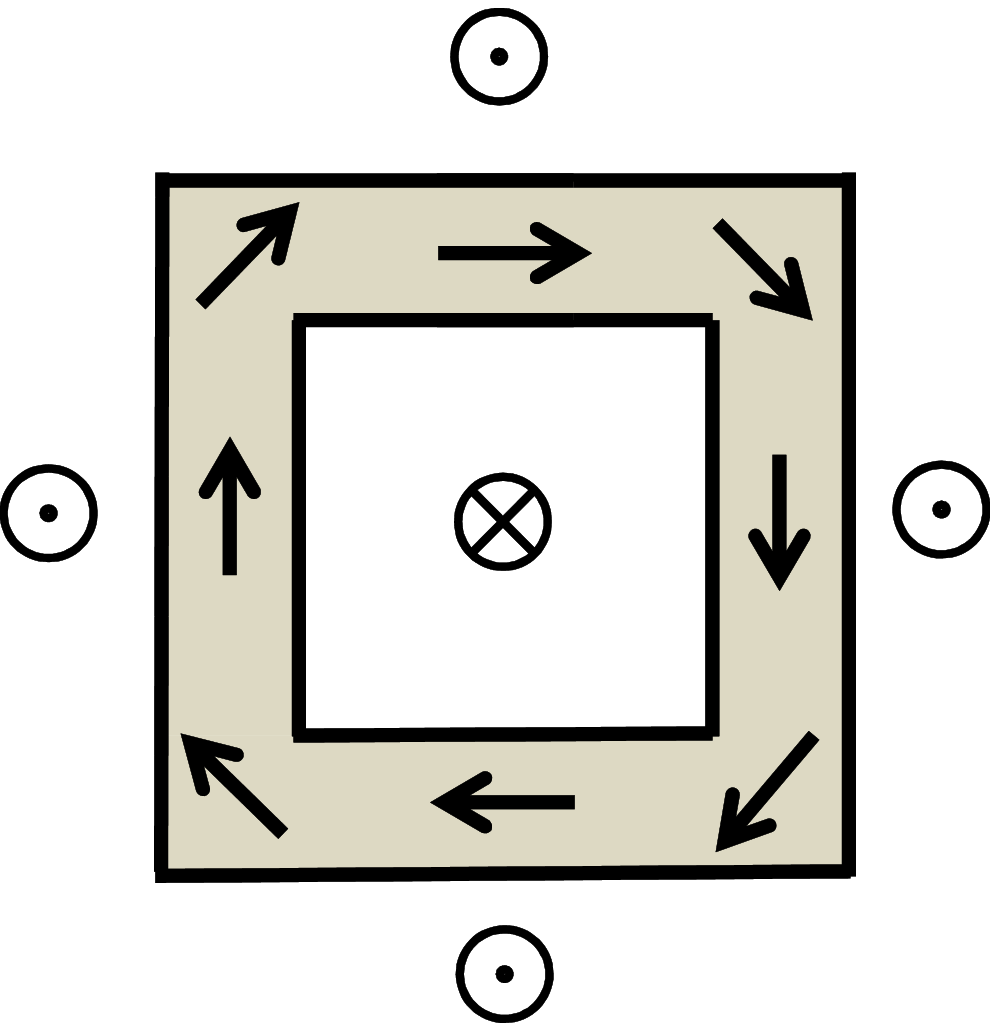}\quad
\includegraphics[width=0.2\linewidth,keepaspectratio]{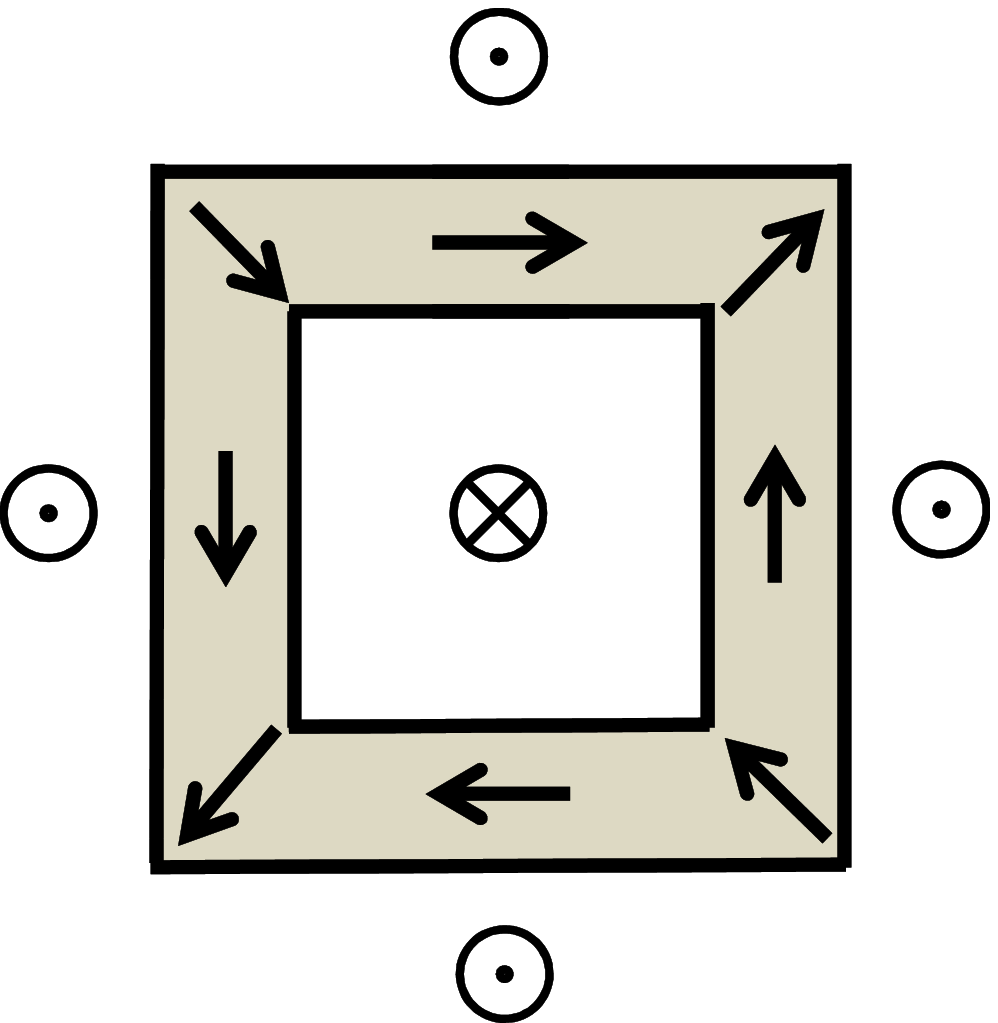}\quad
\includegraphics[width=0.2\linewidth,keepaspectratio]{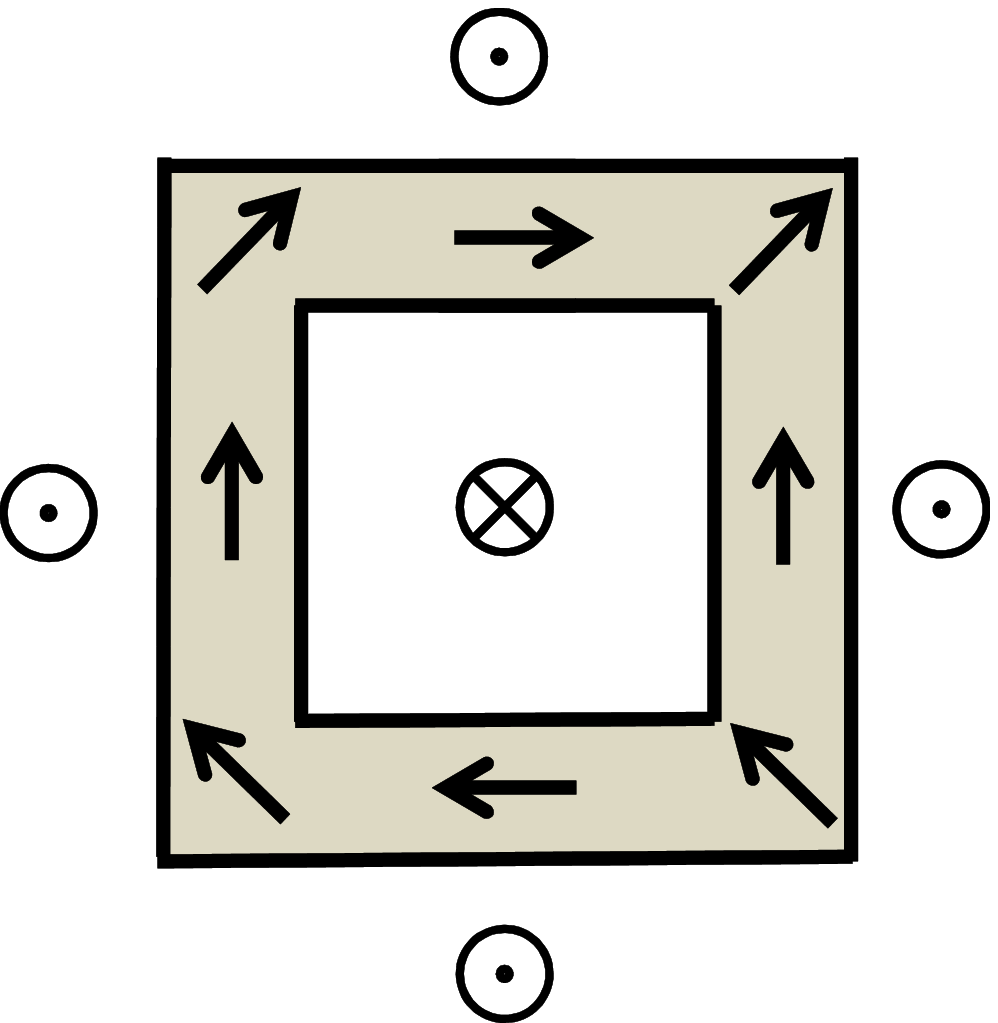}\quad
\includegraphics[width=0.2\linewidth,keepaspectratio]{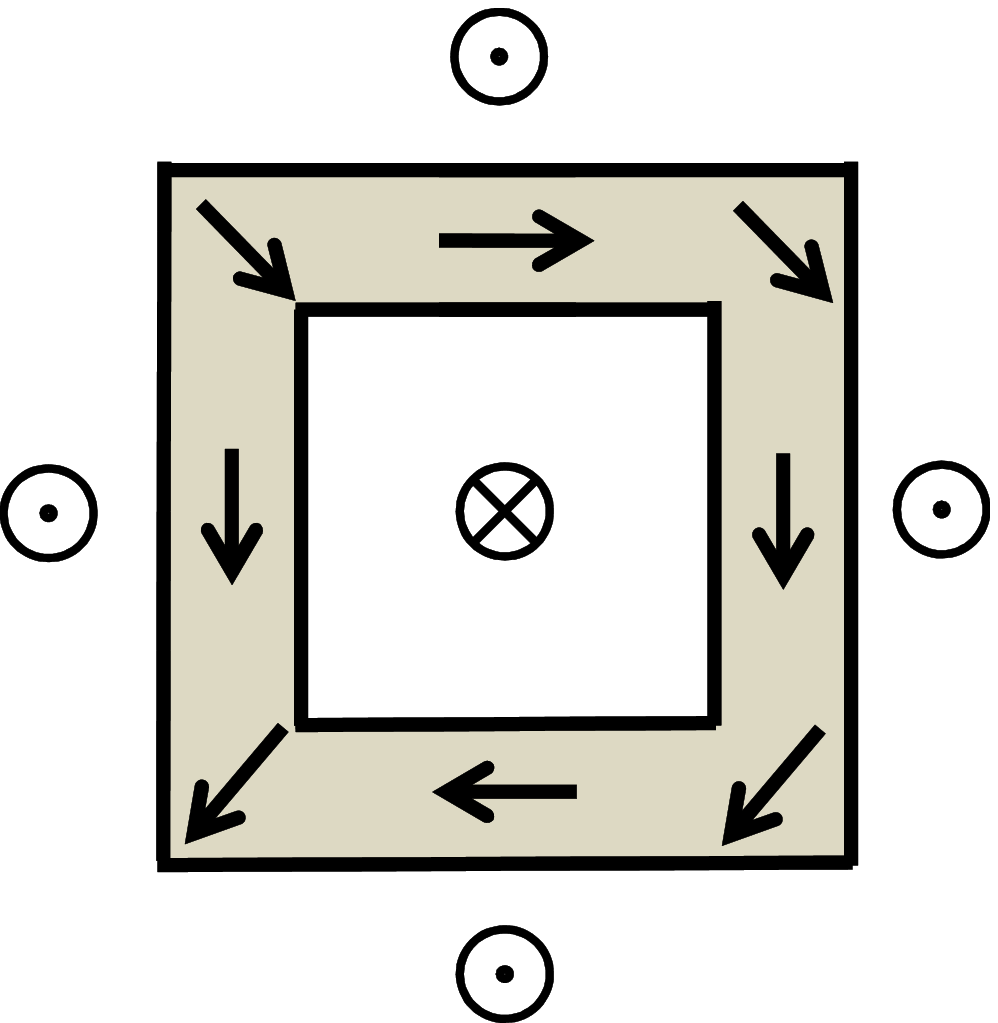}

\hspace{2cm}
(a)\hspace{3cm} (b)\hspace{3cm}
(c)\hspace{3cm} (d)\hspace{3cm} 

\end{center}
\caption{Stable and unstable rings. 
(a,b) Stable rings. The phase winds once 
(the winding number is $\pm 1$) along the rings. 
Total configurations are lumps with 
a non-trivial element, $\pm 1$, of the second homotopy group, $\pi_2$.
(c,d) Unstable rings. 
The phase does not wind  
(the winding number is $0$) along the rings. 
They decay into ground state (up pseudo-spin).
} 
\label{fig:winding} 
\end{figure}

Several holes are created during the entire decaying process.
Let us focus on a pair of two neighboring holes. 
One can find a ring of a domain wall between the holes, 
as shown in Fig.~\ref{fig:winding}.
Here, since there exist two kinds of holes ($\uparrow$ and $\downarrow$), 
there exist four possibilities of the rings,  
(a) $\uparrow\downarrow$, (b) $\downarrow\uparrow$, 
(c) $\uparrow\uparrow$, and (d) $\downarrow\downarrow$  
(Fig.~\ref{fig:winding}). 
Clearly, the rings of types (c) of (d) can decay and 
end up with the vacuum state $\odot$. 
However, the decay of the rings of types 
(a) and (b) is topologically forbidden,  
because of a nontrivial winding of the spin along the rings. 

What are these topologically protected rings of a domain wall?
They are nothing but lumps 
or sigma model instantons \cite{Polyakov:1975yp}, 
which are also known as 2D skyrmions in condensed matter. 
The solutions can be written 
as ($z \equiv x^1+ix^2$)
\begin{equation}
 u = u_0 = \lambda/(z-z_0) \quad \mbox{or} 
\quad u = \bar u_0, \label{eq:lump}
\end{equation} 
as a lump or an anti-lump, 
where $z_0 \in {\bf C}$ represent the positions of the lump   
and $\lambda \in {\bf C}^*$, with  
$|\lambda|$ and $\arg \lambda$
representing the size and $U(1)$ orientation of 
the lump, respectively. 
In fact, 
these configurations can be shown to have a nontrivial winding 
in the second homotopy group, $\pi_2 ({\bf C}P^1)\simeq {\bf Z}$,  
which can be calculated from 
$\displaystyle{ {1\over 2\pi}\int d^2 x \,
  {i (\partial_x u^* \partial_y u - \partial_y u^* \partial_x u )
     \over  (1+|u|^2)^2}}$. 
(a) and (b) belong to, respectively, $+1$ and $-1$ of $\pi_2 ({\bf C}P^1)$.
That is, they are a lump and an anti-lump, respectively.

In the context of magnetism, these configurations 
are referred to as ``bubble domains."

In the presence of the potential term, 
this solution is actually unstable to shrink. 
It can be found that the size $|\lambda|$ tends to zero 
when the configuration in Eq.~(\ref{eq:lump}) 
in substituted in the energy term corresponding to 
the Lagrangian (\ref{eq:Lagrangian}).
In order to avoid this, one can consider 
the ${\bf C}P^1$ model as a low-energy theory for 
the $U(1)$ gauge theory coupled with two complex scalar fields, 
as follows:
\beq
&& {\cal L} = -\1{4 e^2} F_{\mu\nu}F^{\mu\nu} 
 + \1{e^2} (\partial_{\mu} \Sigma)^2 
 + |D_{\mu}\Phi|^2 - V, \non
&& V = {e^2\over 2} (\Phi^\dagger\Phi -v^2)^2 
 + \Phi^\dagger(\Sigma {\bf 1}_2 - M)^2 \Phi
\eeq
with complex scalar fields $\Phi = (\phi^1,\phi^2)^T$ 
and a real scalar field $\Sigma$; here,
$M={\rm diag.}(m_1,m_2)$ with 
$m_1>m_2$ and $m_1-m_2=m$.
This Lagrangian is the bosonic part of 
the ${\cal N}=2$ supersymmetric gauge theories, 
where $v^2$ is called the Fayet-Illiopoulos parameter. 
However, supersymmetry is not essential in our study.

Lumps are replaced with so-called 
semi-local vortices \cite{Vachaspati:1991dz}, 
which also have size moduli 
in the absence of mass $m$. 
In the presence of the mass, however, 
they shrink to the minimum size vortices, {\it i.e.}, 
local (Abrikosov-Nielsen-Olesen) 
vortices \cite{Abrikosov:1956sx}. 

In $d=3+1$ dimensions, 
domain walls have two spatial dimensions 
in their world-volume. 
When a domain-wall pair decay occurs, 
there appear two-dimensional 
holes, which are labeled as 
$\downarrow$ or $\uparrow$ in 
Fig.~\ref{fig:wall-anti-wall-annihilation} (b) or (e), 
respectively.
Along the boundary of these two kinds of holes, 
there appear vortex strings, 
which generally create vortex rings.
This process can be numerically validated \cite{Takeuchi:2011}.
The vortex rings decay into fundamental excitations 
in the end.

\section{Instantons from Monopole--anti-monopole Annihilation}

\begin{figure}[h]
\begin{center}
\includegraphics[width=0.4\linewidth,keepaspectratio]{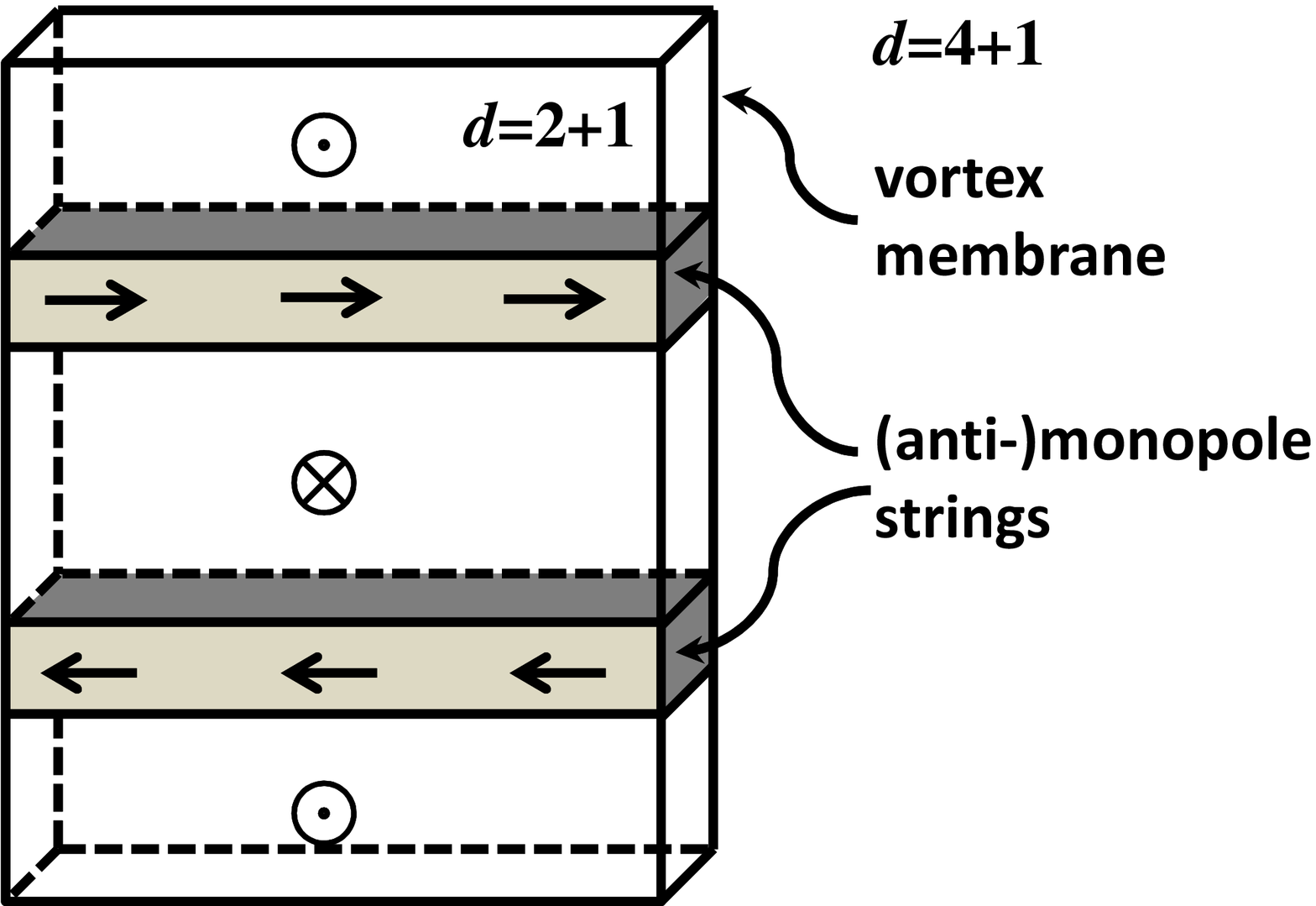}\quad
\includegraphics[width=0.4\linewidth,keepaspectratio]{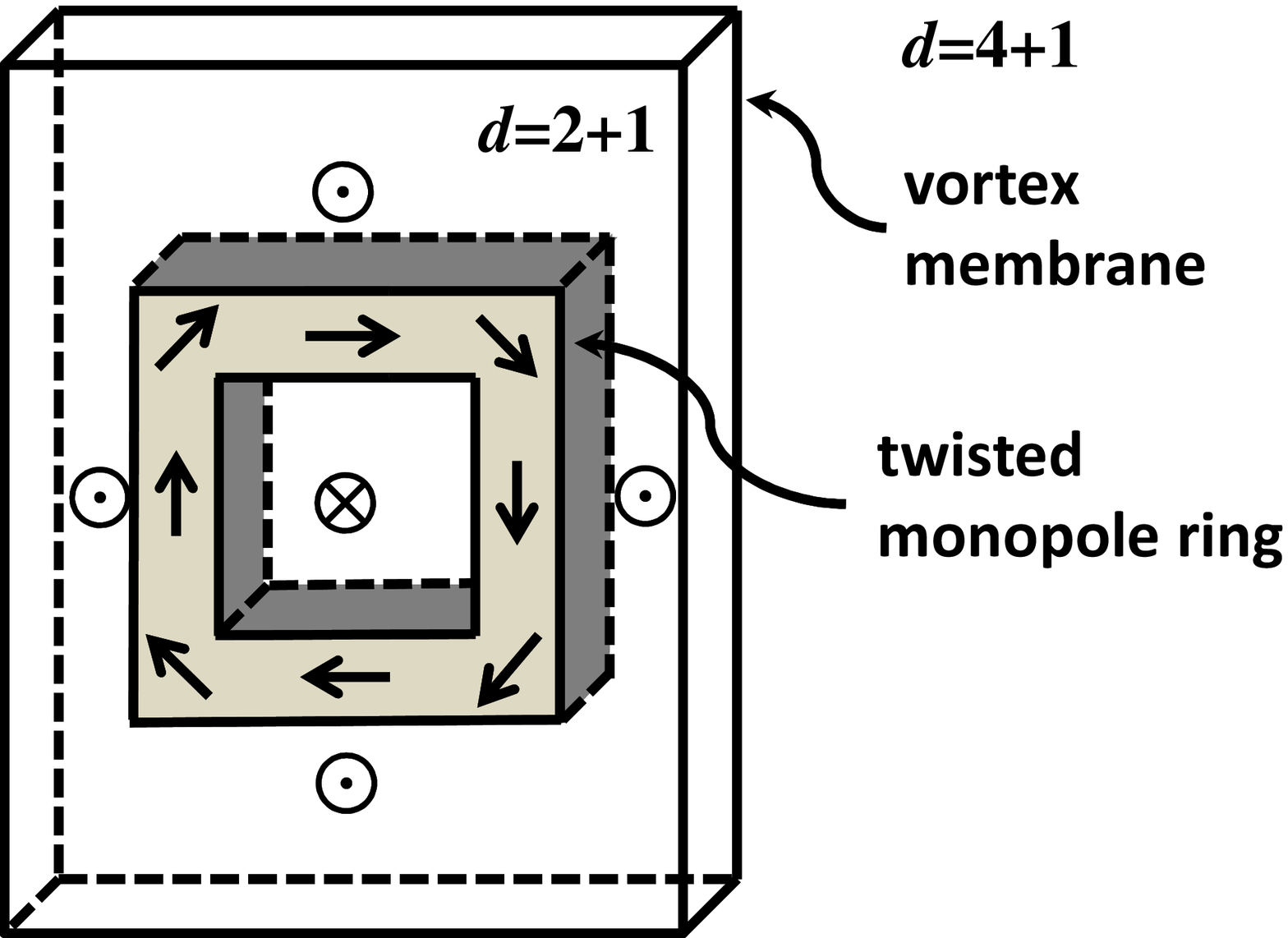}\\
(a)\hs{60} (b)\hs{30}

\caption{(a) A pair of a monopole string and 
an anti-monopole-string inside a non-Abelian vortex. 
(b) A twisted monopole ring inside a non-Abelian vortex. 
\label{fig:monopole-anti-monopole}}
\end{center}
\end{figure}

As the second example, let us consider a pair annihilation of 
't Hooft-Polyakov monopoles.
't Hooft-Polyakov monopoles are 
point-like topological defects in $SU(2)$ Yang-Mills theory, 
coupled with Higgs scalar fields 
in the triplet representation in 
$d=3+1$ dimensions \cite{'tHooft:1974qc}.
In order to create lower-dimensional defects, 
the topological defects to be annihilated require 
at least one world-volume direction. 
Therefore, we consider monopole strings in $d=4+1$ dimensions 
as the minimum set-up.  

The approach involves putting the system into the Higgs phase 
by considering the $U(2)$ gauge theory instead of $SU(2)$, 
where magnetic fluxes from a monopole are squeezed 
into vortices \cite{Tong:2003pz}. 
The Lagrangian in the $d=4+1$ dimensions, 
which we consider, is given by
\beq
&& {\cal L} = - \1{4 g^2}\tr F_{\mu\nu}F^{\mu\nu} 
 + \1{2g^2} \tr (D_{\mu} \Sigma)^2 
  + \tr D_{\mu}H^\dagger D^{\mu}H - V ,\non
&& V = g^2 \tr (H H^\dagger -v^2{\bf 1}_2)^2 
 + \tr \left[H(\Sigma {\bf 1}_2 - M)^2 H^\dagger\right],
 \quad
\eeq
with two complex scalar fields in the 
fundamental representation of $SU(2)$,  
summarized as a two by two complex matrix $H$, 
and with the real matrix scalar fields $\Sigma$
in the adjoint representation. 
The mass matrix is given as
$M={\rm diag.}(m_1,m_2)$, with $m_1>m_2$ and $m_1-m_2=m$.
Again, this Lagrangian 
can be made ${\cal N}=2$ supersymmetric 
by suitably adding fermions;  
however, supersymmetry is not essential in our study.

In the massless limit, $m=0$, 
this model admits a non-Abelian $U(2)$ vortex solution 
$H = {\rm diag.}\, (f(r) e^{i\theta}, v)$ \cite{Hanany:2003hp}.
The solution breaks the vacuum symmetry 
$SU(2)_{\rm C+F}$ into $U(1)$, 
and there appear ${\bf C}P^1 \simeq U(2)_{\rm C+F}/U(1)$ 
Nambu-Goldstone modes.
The low-energy effective theory on 
the $2+1$-dimensional vortex world-volume is the ${\bf C}P^1$ model. 
In the presence of mass, {\it i.e.} $m\neq 0$, 
the $SU(2)_{\rm C+F}$ symmetry is explicitly broken, 
inducing the same type of mass matrix in the $d=2+1$ 
vortex effective theory; 
\beq
&& {\cal L}_{\rm vort.eff.}
=  2 \pi v^2 |z_0|^2 + {4\pi \over g^2} \left[ 
 {\partial_{\mu} u^* \partial^{\mu} u - m^2 |u|^2 
  \over (1 + |u|^2)^2} \right] .
\eeq

Let us construct a domain wall in this effective theory.
The domain wall tension is $4\pi m/g^2$, 
which coincides with the monopole mass.
It is equivalent to the monopole charge 
because the monopole is a BPS state. 
Therefore, the wall in the vortex theory 
is nothing but a monopole string from  
the bulk point of view \cite{Tong:2003pz}. 

We then consider a pair of a monopole string and 
an anti-monopole string in the vortex world-volume, 
as in Fig.~\ref{fig:monopole-anti-monopole}~(a).
The result in the last section immediately 
implies that 
after a monopole pair annihilation,  
we obtain lumps \cite{Polyakov:1975yp} 
in the vortex world-volume, as shown in 
 Fig.~\ref{fig:monopole-anti-monopole}~(b).
As shown in \cite{Eto:2004rz},
lumps inside a non-Abelian vortex are nothing but 
Yang-Mills instantons \cite{Belavin:1975fg},
which are point-like topological objects from 
the $d=4+1$ bulk point of view. 
This can also be inferred from the lump energy, 
which coincides with the instanton energy.

Finally, we send the FI parameter $v^2$ to zero 
with the system going back to the Coulomb phase.
Then, without the help of non-Abelian vortices, 
the monopole-string pair annihilates,  
resulting in the creation of (anti-)instantons.\footnote{ 
In the Coulomb phase, 
monopole strings can move in their three codimensions,  
but in the Higgs phase they can move in only 
one codimension inside a vortex. 
Therefore, monopole strings generally collide at an angle. 
In this case, they reconnect with each other, 
resulting in (un)twisted monopole rings, 
as discussed in this paper.
We can show that 
such a reconnection always occurs owing to 
from the fact that when two monopoles collide head on 
in d=3+1 dimensions, 
they scatter at a 90$^\circ$ angle.
}
We also have found that a twisted monopole ring 
is an instanton.

\section{Summary and discussion}
In conclusion, when a pair of a defect and an anti-defect 
with opposite $U(1)$ moduli annihilates, 
there appear (anti-)defects with 
dimensions less than those of the original defects by one.
We have demonstrated this phenomenon in two examples:
1) when a domain wall and an anti-domain wall annihilate 
in 2+1 dimensions, there appear (anti-)vortices, 
and 2) when a monopole string and an anti-monopole string 
annihilate in 4+1 dimensions, there appear 
(anti-)Yang-Mills instantons. 
We have found that twisted domain wall rings and 
twisted monopole rings 
are vortices and instantons, respectively.

Another example is that of 
superconducting cosmic strings 
that have a $U(1)$ zero mode \cite{Witten:1984eb}.
When a pair of superconducting strings annihilates, 
with opposite $U(1)$ moduli, 
there can appear a vorton, 
{\it i.e.} a twisted vortex ring. 
It is an open question if this vorton can be identified with a monopole 
in some setting. 

Studies on defect-anti-defect annihilations in this paper 
have been restricted to semi-classical analysis.
Quantum mechanical decay of 
metastable topological defects 
was studied before \cite{Preskill:1992ck}.
The extension to quantum mechanical decay of 
defect--anti-defect annihilation remains as an important 
future work.

{\bf Acknowledgements}

This work is supported in part by 
Grant-in Aid for Scientific Research (No. 23740198) 
and by the ``Topological Quantum Phenomena'' 
Grant-in Aid for Scientific Research 
on Innovative Areas (No. 23103515)  
from the Ministry of Education, Culture, Sports, Science and Technology 
(MEXT) of Japan.

\end{document}